\documentclass[letterpaper, 10 pt, conference,twocolumn]{ieeeconf}
\IEEEoverridecommandlockouts                              
\usepackage{cite}

\usepackage{array,pdfsync,enumerate,tikz,enumerate}
\usepackage[dvips]{epsfig}
\usepackage{amssymb,amsmath}
\usepackage{caption}
\usepackage{subcaption}
\usepackage{color}
\definecolor{linkcol}{rgb}{0,0.0,1.0}
\definecolor{citecol}{rgb}{0.0,0.6,0.0}

\IEEEoverridecommandlockouts
\usepackage{cite}
\usepackage{amsmath,amssymb,amsfonts}
\usepackage{graphicx}
\usepackage{textcomp}

\usepackage{hyperref}
\hypersetup{colorlinks,breaklinks,
	linkcolor=linkcol,urlcolor=citecol,
	anchorcolor=linkcol,citecolor=citecol,bookmarks=true}

\usepackage{caption}
\usepackage{subcaption}

\newtheorem{theorem}{Theorem}[section]
\newtheorem{proposition}[theorem]{Proposition}
\newtheorem{lemma}[theorem]{Lemma}
\newtheorem{remark}[theorem]{Remark}
\usepackage{graphicx}      
\usepackage{cite}
\begin{document}
	
	\title{Chetaev Instability Framework for Kinetostatic Compliance-Based Protein Unfolding}
	
	\author{Alireza Mohammadi and Mark W. Spong
		\thanks{This work is supported by the National Science Foundation (NSF) under Grants 2153744 and 2153901. A. Mohammadi is with the Department of Electrical \& Computer Engineering, University of 
			Michigan-Dearborn, Dearborn, MI 48128 USA. Mark W. Spong is with the
			University of Texas at Dallas, Richardson, TX 75080 USA. 
			Emails: {\tt\small \{amohmmad@umich.edu$^{\ast}$, mspong@utdallas.edu\}}, $^{\ast}$Corresponding Author: A. Mohammadi}%
	}

\maketitle		
		
		\begin{abstract}                
			  \textcolor{black}{Understanding the process of protein unfolding plays a
			  	crucial role in various applications such as design of folding-based protein engines.  Using the well-established kinetostatic compliance (KCM)-based method for modeling of protein conformation dynamics and a recent nonlinear control theoretic approach to KCM-based protein
				folding, this paper formulates protein unfolding as a destabilizing control analysis/synthesis problem. In light of this formulation, it is shown that the Chetaev instability framework can be used to investigate the KCM-based unfolding dynamics.  \textcolor{black}{In particular, a Chetaev function for analysis of unfolding dynamics under the effect of optical tweezers and a class of control Chetaev functions for synthesizing control inputs that elongate protein  strands from their folded conformations are presented. Based on the presented control Chetaev function, an unfolding input is derived from the  Artstein-Sontag universal formula and the results are compared against optical tweezer-based unfolding.}}
		\end{abstract}
		


\renewcommand\floatpagefraction{.9}
\renewcommand\dblfloatpagefraction{.9} 
\renewcommand\topfraction{.9}
\renewcommand\dbltopfraction{.9} 
\renewcommand\bottomfraction{.9}
\renewcommand\textfraction{.1}   
\setcounter{totalnumber}{50}
\setcounter{topnumber}{50}
\setcounter{bottomnumber}{50}
\newenvironment{psmallmatrix}
{\left(\begin{smallmatrix}}
	{\end{smallmatrix}\right)}
%
    \section{Introduction}
Protein molecules make conformational transitions in-between two or more native conformations to perform a plethora of crucial biological functions~\cite{finkelstein2016protein}. These transitions usually involve two distinct processes of folding and unfolding. During the folding/unfolding process, a polypeptide chain folds/unfolds into/from a biologically active molecule (\textcolor{black}{see Supplementary Material for an energy landscape schematic of folding/unfolding}).  In addition to its role for understanding the protein structures, unfolding is used for engineering materials such as hydrogels and anti-viral drugs as well as designing protein-based nanomachines~\cite{bergasa2020interdiction,fang2013forced,fu2019harvesting}. 

Algorithmic prediction of pathways through which a folded protein molecule unfolds to its denatured  state has been among the main tools for investigating the  biochemical process of unfolding. The AI-based approaches to predicting the folded structure of proteins such as Google AlphaFold~\cite{alquraishi2020watershed}, which  are rooted in pattern recognition, cannot address issues such as folding/unfolding pathway prediction. \textcolor{black}{Indeed, AlphaFold's main task has been to predict the most likely structures of folded  proteins  while ignoring the kinetics and stability of folding/unfolding processes (see, e.g.,~\cite{fersht2021alphafold} for further details).} Alternatively, dynamics-based approaches rooted in physical first-principles provide a more preferred way for predicting the folding/unfolding pathways~\cite{heo2019driven}. 


To address the high computational burden of physics-based approaches, the framework of kinetostatic compliance method (KCM), which  is based on modeling protein molecules as mechanisms with a large number of rigid nano-linkages that fold under the nonlinear effect of electrostatic and van der Waals interatomic forces, has been developed in the literature~\cite{alvarado2003rotational,kazerounian2005nano,kazerounian2005protofold,tavousi2015protofold,Tavousi2016}. The KCM framework has been successful in investigating the effect of hydrogen bond formation on protein kinematic mobility~\cite{shahbazi2010hydrogen} and synthesizing molecular nano-linkages~\cite{tavousi2016synthesizing,chorsi2021kinematic}. Furthermore, as demonstrated by~\cite{mohammadi2021quadratic}, this framework lends itself to a nonlinear control theoretic interpretation, where entropy-loss constraints can be encoded in KCM-based folding via proper quadratic optimization-based nonlinear control techniques.  Despite its success for investigating the structure of proteins, the KCM framework except for a simulation study in~\cite{su2007comparison} has not been systematically utilized for studying unfolding.    

In this paper, we formulate the KCM-based protein unfolding as a destabilizing control analysis/synthesis problem by utilizing the Chetaev instability  framework. \textcolor{black}{A systemic development of Chetaev functions (CFs) and control Chetaev functions (CCFs), which are rooted in the seminal work of N. G. Chetaev~\cite{chetaev1961stability}, has been recently initiated by Efimov and collaborators~\cite{efimov2009oscillatority,efimov2011oscillating,efimov2014necessary} and Kellett and collaborators~\cite{braun2018complete}. CFs and CCFs are not only useful for studying instability of nonlinear control systems but also for synthesis of destabilizing control inputs to induce oscillations via static feedback and safety critical control applications.} CFs and CCFs in this paper are employed to investigate the instability properties of the folded protein conformations under study. 

\noindent{\textbf{Contributions of the paper.}} This paper has the following contributions. First, this paper bridges the two fields of protein unfolding in biochemistry and the Chetaev instability framework  in nonlinear control theory.  Second, this paper provides a nonlinear control theoretic foundation for single-molecule force-based protein denaturation. The single-molecule protein unfolding using devices such as optical tweezers~\cite{bustamante2020single} and atomic force microscopes (AFMs)~\cite{ragazzon2018model} has proven to be useful in nano-/molecular-robotics~\cite{thammawongsa2012nanorobot,wen2019nanorobotic}.  Third, this paper adds to the body of knowledge on KCM-based modeling for further investigation of structural properties of protein molecules.  

The rest of this paper is organized as follows. First, we review the KCM framework for modeling 
protein molecules, the control theoretic interpretation of it, and \textcolor{black}{some basics from the Chetaev instability theory} in  Section~\ref{sec:dynmodel}. Thereafter, \textcolor{black}{we model the dynamics of protein unfolding under optical tweezers} in Section~\ref{sec:Sol}. Next,  \textcolor{black}{a CF for analysis of optical tweezer-based unfolding and a class of CCFs for synthesizing unfolding control inputs, which elongate protein molecules, are presented  in  Section~\ref{sec:ContProb}.} \textcolor{black}{After presenting the simulation results that compare optical tweezer-based unfolding against a synthesized  Artstein-Sontag destabilizing feedback} in Section~\ref{sec:sims}, we  conclude the paper in Section~\ref{sec:conc}.   

\noindent{\textbf{Notation.}} We let $\mathbb{R}_{+}$ denote the set of all non-negative 
real numbers. We let $(\cdot)^\top$ denote the vector/matrix transpose operator. Given $\mathbf{x}\in\mathbb{R}^M$, we let $|\mathbf{x}|:=\sqrt{\mathbf{x}^{\top}\mathbf{x}}$
denote the Euclidean norm of $\mathbf{x}$. Given $\varepsilon\in \mathbb{R}_{+}$, we let $\mathcal{B}_{\mathbf{x}}(\varepsilon)$ denote the open ball centered at $\mathbf{x}$ with radius $\varepsilon$. Given two vectors $\mathbf{v}_1,\, \mathbf{v}_2 \in \mathbb{R}^3$, we let $\mathbf{v}_1 \times \mathbf{v}_2$ denote their cross product. 

    \section{Preliminaries}
\label{sec:dynmodel} 
%
%
%
\begin{figure}[t]
	\centering
	\includegraphics[width=0.45\textwidth]{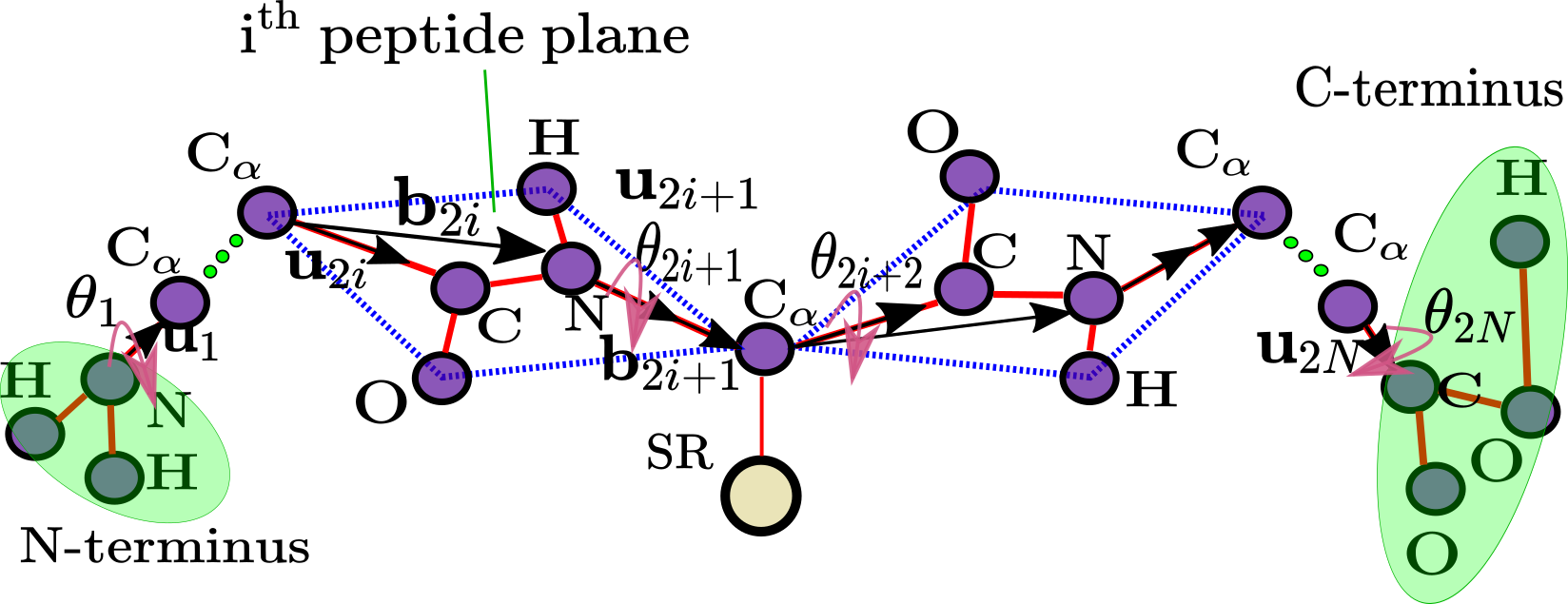}
	\caption{\small The protein molecule kinematic structure.}
	\vspace{-4.5ex}
	\label{fig:protBasic}
\end{figure}
\subsection{Kinematic linkage representation of protein molecules}
\label{subsec:kinLinkage}
In this section, we review the KCM-based modeling of the main chain \emph{in vacuo}  (see~\cite{tavousi2015protofold,Tavousi2016} for further details). Proteins essentially consist of peptide planes that are rigidly joined together in a long chain (see Figure~\ref{fig:protBasic}). Central carbon atoms denoted by $\text{C}_\alpha$ play the role of hinges in-between the amino acids.  The red line segments in Figure~\ref{fig:protBasic} represent covalent atomic bonds. The first $\text{C}_\alpha$  is connected to the N-terminus, and one other peptide plane. Similarly, the last $\text{C}_\alpha$ hinges to the C-terminus, and one other peptide plane.

The backbone conformation of protein molecules can be described by  using a collection of 
bond lengths and two sets of rotation angles around $\text{N}-\text{C}_\alpha$ and 
$\text{C}_\alpha-\text{C}$, which are known as the dihedral angles. Hence, the vector of dihedral angles 
$
\pmb{\theta}= \big[\theta_1,\cdots,\theta_{2N}]^\top \in \mathbb{R}^{2N}
$
represents the protein configuration  with $N-1$ peptide planes. Along the rotation axis corresponding to  each DOF of the protein molecule, a unit vector denoted by $\mathbf{u}_j$, $1\leq j \leq 2N$, can be considered. 

In addition to the unit vectors $\mathbf{u}_j$, Kazerounian and collaborators~\cite{kazerounian2005nano,kazerounian2005protofold,tavousi2015protofold}  utilize the so-called \textbf{body vectors} to represent the spatial orientation  of the peptide planes. These body vectors are denoted by $\mathbf{b}_{j}$, $1\leq j \leq 2N$, and  describe the relative position of any two atoms in the chain.   Using the vectors $\mathbf{u}_j$ and $\mathbf{b}_{j}$, the conformation of 
a protein molecule can be completely described. In particular, after designating the zero position conformation with $\pmb{\theta}=\mathbf{0}$ corresponding to the biological reference position of the chain (see~\cite{kazerounian2005nano}), the transformations 
%
\begin{equation}
\mathbf{u}_j(\pmb{\theta}) = \Xi(\pmb{\theta},\mathbf{u}_j^0) \mathbf{u}_j^0  ,\, 
\mathbf{b}_j(\pmb{\theta}) = \Xi(\pmb{\theta},\mathbf{u}_j^0) \mathbf{b}_j^0,
\label{eq:kineProtu}
\end{equation}
where 
\begin{equation}
\Xi(\pmb{\theta},\mathbf{u}_j^0)=\prod_{r=1}^{j} R(\theta_j,\mathbf{u}_j^0), 
\label{eq:kineProtMat}
\end{equation}
describe the molecule kinematic configuration. In~\eqref{eq:kineProtMat}, the 
 matrix $R(\theta_j,\mathbf{u}_j^0)\in \text{SO}(3)$ represents the rotation about the unit vector 
$\mathbf{u}_j^0$ with angle $\theta_j$. Having obtained the body 
vectors $\mathbf{b}_j(\pmb{\theta})$ from~\eqref{eq:kineProtu} and under the assumption that the N-terminus nitrogen atom is fixed at the 
origin, the coordinates of the backbone chain atoms in the $k$-th peptide plane are found from
$
\mathbf{r}_i(\pmb{\theta}) = \sum_{j=1}^{i}  \mathbf{b}_j(\pmb{\theta})
$.
\subsection{Control theoretic interpretation of KCM-based dynamics}
The KCM framework in~\cite{kazerounian2005protofold,tavousi2015protofold,Tavousi2016} builds upon the experimental fact that the inertial forces during the folding of a protein chain can be neglected in the presence of the interatomic forces (see, e.g.,~\cite{adolf1991brownian,kazerounian2005protofold}).  Instead, the dihedral angles rotate under the kinetostatic influence of these forces.

 To present the KCM-based protein dynamics, let us consider a polypeptide chain with $N-1$ peptide planes and conformation vector $\pmb{\theta}\in \mathbb{R}^{2N}$.  The  molecule aggregated free energy is given by (see~\cite{Tavousi2016} for detailed derivations)
 \begin{equation}\label{eq:freePot}
 \mathcal{G}(\pmb{\theta}) := \mathcal{G}^{\text{elec}}(\pmb{\theta}) + \mathcal{G}^{\text{vdw}}(\pmb{\theta}),
 \end{equation}
 where $\mathcal{G}^{\text{elec}}(\pmb{\theta})$ and  $\mathcal{G}^{\text{vdw}}(\pmb{\theta})$ represent the 
 the electrostatic and van der Waals potential energies, respectively.  
 
 
The resultant forces and torques on each of the $N-1$ rigid peptide links can be computed from the molecule free energy in~\eqref{eq:freePot} and the protein chain kinematics. These forces and torques can then be appended in the generalized force vector $\mathcal{F}(\pmb{\theta})\in \mathbb{R}^{6N}$. Following the steps of derivation in~\cite{tavousi2015protofold}, one can utilize a proper mapping to relate $\mathcal{F}(\pmb{\theta})$ to the equivalent torque vector $\pmb{\tau}(\pmb{\theta})\in \mathbb{R}^{2N}$ acting on the dihedral angles at conformation $\pmb{\theta}$, which is given by 
\begin{equation}\label{eq:tau_KCM}
\pmb{\tau}(\pmb{\theta}) = \mathcal{J}^\top(\pmb{\theta}) \mathcal{F}(\pmb{\theta}).
\end{equation}

In~\eqref{eq:tau_KCM}, the matrix  $\mathcal{J}(\pmb{\theta})\in \mathbb{R}^{6(2N)\times 2N}$ 
represents the Jacobian of the protein chain at conformation $\pmb{\theta}$. The vector  $\mathcal{F}(\pmb{\theta})$  is due to the torques and forces, which arise from the interatomic electrostatic and van der Waals effects, at conformation $\pmb{\theta}$. As it is demonstrated in~\cite{tavousi2015protofold,kazerounian2005protofold}, for a protein molecule with $N-1$ peptide planes,  the matrix $\mathcal{J}^\top(\pmb{\theta})\in \, \mathbb{R}^{2N \times 6(2N)}$ has the structure 
\begin{equation}
\mathcal{J}^\top(\pmb{\theta})  = \begin{bmatrix} \mathbf{J}_1^\top(\pmb{\theta}) & \mathbf{J}_1^\top(\pmb{\theta}) & \cdots  & \mathbf{J}_1^\top(\pmb{\theta}) \\
\textbf{0} & \mathbf{J}_2^\top(\pmb{\theta}) & \cdots  & \mathbf{J}_2^\top(\pmb{\theta}) \\
\vdots & \vdots & \vdots & \vdots \\
\textbf{0} & \textbf{0} & \cdots & \mathbf{J}^\top_{2N}(\pmb{\theta})
\end{bmatrix},
\label{eq:overallJacob}  
\end{equation}
where 
\begin{equation}
\mathbf{J}_k(\pmb{\theta}) = \begin{bmatrix} \mathbf{u}_k(\pmb{\theta}) \\ - \mathbf{u}_k(\pmb{\theta}) \times \mathbf{r}_k(\pmb{\theta}) \end{bmatrix} \in \, \mathbb{R}^{6},\, 1 \leq k \leq 2N, 
\label{eq:localJacob}  
\end{equation}
in which $\mathbf{u}_k(\pmb{\theta})$ denotes the unit vector introduced in Section~\ref{subsec:kinLinkage} and $\mathbf{r}_k(\pmb{\theta})$ connects the N-terminus to either $\text{C}_\alpha$ atom (even $k$) or $N$ atom (odd $k$) on the $k$-th peptide plane.  

Using the torque vector in~\eqref{eq:tau_KCM} and starting from an initial conformation 
 $\pmb{\theta}_0$, the KCM-based iteration is given by 
%
$\pmb{\theta}_{i+1} =  \pmb{\theta}_{i} + h \textbf{f}_\text{KCM} \big(\pmb{\tau}(\pmb{\theta}_i)\big),$ 
%
\noindent where $i$ is a non-negative integer, $h$ is a positive real constant that tunes the maximum dihedral angle rotation magnitude in each step, and $\textbf{f}_\text{KCM}(\cdot)$ is a proper mapping, which can be chosen to be the identity map in its simplest form  (see~\cite{tavousi2015protofold} for further details).  The successive iteration of the KCM-based folding is performed until all of the kinetostatic torques converge to a minimum corresponding to a local minimum of the free energy. Furthermore, $\mathcal{J}^\top(\pmb{\theta}) \mathcal{F}(\pmb{\theta})$ is a locally  Lipschitz continuous function in a neighborhood of a given folded conformation.

In~\cite{mohammadi2021quadratic}, it has been demonstrated that the KCM-based iteration can be interpreted as a control synthesis problem. In particular, the work in~\cite{mohammadi2021quadratic} considered the following nonlinear control-affine system
\begin{equation}\label{eq:KCM_ode}
	\dot{\pmb{\theta}} = \mathcal{J}^\top(\pmb{\theta}) \mathcal{F}(\pmb{\theta}) + \mathbf{u}_{c},
\end{equation}
\hspace{-1ex} where the Jacobian matrix $\mathcal{J}(\pmb{\theta})$ and the generalized force vector $\mathcal{F}(\pmb{\theta})$ 
are the same as in~\eqref{eq:tau_KCM}.  It was demonstrated that forward Euler discretization of the closed-loop dynamics in~\eqref{eq:KCM_ode} under a proper closed-loop feedback control input $ \mathbf{u}_{c}(\pmb{\theta})$ restores the KCM-based iteration for protein folding. 
\vspace{-0ex}
\textcolor{black}{
\subsection{Chetaev instability theory preliminaries}}
\label{subsec:chetaevPrelim}
\textcolor{black}{Consider the nonlinear dynamical system 
\begin{equation}
 \dot{\mathbf{x}}= \mathbf{F}(\mathbf{x}),
 \label{eq:nonlinSys}
\end{equation}
where $\mathbf{F}:\mathbb{R}^n \to \mathbb{R}^n$ is a locally Lipschitz continuous function.  We say that the equilibrium $\mathbf{x}^\ast \in \mathbb{R}^n$ is unstable for~\eqref{eq:nonlinSys} if for any $\delta>0$ there exists $\mathbf{x}_0 \in \mathcal{B}_{\mathbf{x}^\ast}(\delta)$ such that 
$\mathbf{x}(T_{\mathbf{x}_0},\mathbf{x}_0)\notin \mathcal{B}_{\mathbf{x}^\ast}(\delta)$ for some $T_{\mathbf{x}_0} \in \mathbb{R}_+$, where the solution of~\eqref{eq:nonlinSys} with  $\mathbf{x}(0)=\mathbf{x}_0$ is given by $\mathbf{x}(\cdot,\mathbf{x}_0)$ on the solution maximal interval of existence. 
\begin{theorem}[\cite{chetaev1961stability}]
Consider the system in~\eqref{eq:nonlinSys} and the equilibrium $\mathbf{x}^\ast$. Let $V: \mathbb{R}^n \to \mathbb{R}$ be a locally Lipschitz continuous function such that  $V(\mathbf{x}^\ast)=0$. Suppose that there exists $\epsilon_0>0$ such that $\mathcal{V}^+ \cap \mathcal{B}_{\mathbf{x}^\ast}(\epsilon)\neq \emptyset$ for any $\epsilon\in (0,\epsilon_0]$, where $\mathcal{V}^+:=\{\mathbf{x}\in\mathcal{B}_{\mathbf{x}^\ast}(\epsilon): V(\mathbf{x})>0\}$. If the lower Dini derivative of $V$ along $\mathbf{F}$ satisfies $D^{-}V(\mathbf{x})\mathbf{F}(\mathbf{x})>0$ for all $\mathbf{x}\in\mathcal{V}^+$, then the equilibrium $\mathbf{x}^\ast$ is unstable for~\eqref{eq:nonlinSys}. 
\label{thm:cheta}	
\vspace{-0.25ex}
\end{theorem}} 

\textcolor{black}{A function $V(\cdot)$ that satisfies the conditions in Proposition~\ref{thm:cheta} is called a CF for system~\eqref{eq:nonlinSys}. While the classical theorem by Chetaev~\cite{chetaev1961stability} only provided sufficient conditions for instability, Efimov and collaborators provided the necessary part of the Chetaev's theorem in~\cite{efimov2014necessary}. Furthermore, they developed 
the CCF concept as a counterpart of the Control Lyapunov function (CLF) theory ~\cite{efimov2014necessary}. Kellett and collaborators~\cite{braun2018complete} extended the work in~\cite{efimov2014necessary} to differential inclusions. See Supplementary Material for the required background on CCFs.} 
    \section{KCM-Based Modeling of Protein Unfolding under Optical Tweezers}
\label{sec:Sol}

In this section we present the dynamics of protein unfolding under the effect of an optical tweezer. \textcolor{black}{We remark that the derivation of KCM-based dynamical model of protein unfolding under optical tweezers has not been presented elsewhere in the literature.} 

Optical tweezers can be utilized for unfolding protein molecules by chemically attaching a latex bead (microsphere) to the protein under study, where the bead can be trapped  and used as a handle for stretching the molecule. \textcolor{black}{The reader is referred to Supplementary Material for further details.}  The optical tweezer tension force on the molecule strand can be modeled by the Hookean force~\cite{bustamante2020single}
\begin{equation}
\textbf{F}_\text{twz}(\pmb{\theta}) = \kappa(\pmb{\theta}) \mathbf{x}_\text{twz}, 
\label{eq:optTwzForce}
\end{equation}
\noindent where $\mathbf{x}_\text{twz}$ represents the displacement of the microsphere from the center of the beam and $\kappa(\pmb{\theta})$ is the optical trap stiffness modulated by the protein conformation using a proper technique such as  linear polarization~\cite{schonbrun2008spring,li2015tracking}. In particular, we assume that the trap stiffness can be modulated as a smooth function of $\pmb{\theta}$ in a way that at a given folded conformation $\pmb{\theta}^\ast$, $\kappa(\pmb{\theta}^\ast)=0$.  
\begin{remark}
	For a given positive integer $m$ and a positive constant $\kappa_0$, a candidate optical trap modulated stiffness is 
	\begin{equation}
	\kappa(\pmb{\theta}) = \kappa_0 |\mathbf{r}_\text{NC}(\pmb{\theta}^\ast) - \mathbf{r}_\text{NC}(\pmb{\theta})|^m,
	\label{eq:optTrapStiff2}
	\end{equation}
	\label{rem:modul}
	\hspace{-1ex}where at any conformation $\pmb{\theta}$, the vector $\mathbf{r}_\text{NC}(\pmb{\theta})\in \mathbb{R}^3$ 
	connects the nitrogen atom in the $\text{N}$-terminus to the carbon atom in the $\text{C}$-terminus  (see Figure~\ref{fig:protBasic}).
\end{remark}
\vspace{0ex}
\hspace{1ex} Let us consider the optical tweezer force in~\eqref{eq:optTwzForce} and a protein molecule chain with $N-1$ peptide planes. Without loss of generality, we assume that this force is being exerted on the C-terminus of the protein molecule. Accordingly, the resultant torque applied to the last link $\text{AA}_{\text{last}}$ of the molecule at conformation $\pmb{\theta}$, is given by
$
\mathbf{T}_\text{twz}(\pmb{\theta}) = \sum_{a_i \in \text{AA}_{\text{last}}} \mathbf{r}_i(\pmb{\theta}) \times \kappa(\pmb{\theta})  \mathbf{x}_\text{twz}, 
$
 where $\mathbf{r}_i(\pmb{\theta})$ denotes the position of atom $a_i$, which is the $i$-th atom located in peptide plane $\text{AA}_{\text{last}}$. Therefore, the force-torque couple due to the effect of the optical tweezer is  
$
\mathbf{TF}_\text{twz}(\pmb{\theta}):= \begin{bmatrix} \mathbf{T}_\text{twz}(\pmb{\theta})^\top ,  \mathbf{F}_\text{twz}(\pmb{\theta})^\top \end{bmatrix}^\top \in \mathbb{R}^6. 
$

Consequently, the optical tweezer-based force-torque couple, which is exerted on the last peptide plane of the protein molecule, gives rise to the equivalent joint torque vector 
\begin{equation}
\mathbf{u}_\text{unfold}(\pmb{\theta}) = \mathcal{J}^\top(\pmb{\theta}) \mathcal{F}_\text{twz}(\pmb{\theta}), 
\label{eq:unfoldTorque}
\end{equation}
where $\mathcal{F}_\text{twz}(\pmb{\theta})$ is the generalized force vector 
\begin{equation}
\mathcal{F}_\text{twz}(\pmb{\theta}) = \begin{bmatrix} \mathbf{0}^\top , \cdots , \mathbf{0}^\top , \mathbf{TF}_\text{twz}(\pmb{\theta})^\top\end{bmatrix}^\top \in\; \mathbb{R}^{6(2N)},
\label{eq:genForce}
\end{equation}
%
which gets mapped through~\eqref{eq:unfoldTorque} to the vector $\mathbf{u}_\text{unfold}(\pmb{\theta})\in \mathbb{R}^{2N}$ affecting the dihedral angles in the molecule. 
    \section{Destabilizing control analysis/synthesis for studying the KCM-based protein unfolding}
\label{sec:ContProb}

In this section we first formulate the KCM-based protein unfolding as a 
destabilizing control  analysis/synthesis problem for~\eqref{eq:KCM_ode}. 
\textcolor{black}{Next, we present a
CF for analysis of a folded conformation instability under optical tweezers. Thereafter, we introduce a class of CCFs for synthesis of destabilizing inputs that elongate protein strands from their folded conformations.}
\subsection{\textcolor{black}{Control problem formulation and a trivial solution}}
\label{subsec:pucp}
It is known that the KCM-based iteration when initiated in a vicinity of a stable conformation $\pmb{\theta}^\ast$ converges to it~\cite{kazerounian2005protofold,tavousi2015protofold}. The conformation $\pmb{\theta}^\ast$ corresponds to a local minimum of the aggregated free energy function $\mathcal{G}(\cdot)$ in~\eqref{eq:freePot}, where the equivalent torque in~\eqref{eq:tau_KCM}  becomes   $\pmb{\tau}(\pmb{\theta}^\ast)=\mathcal{J}^\top(\pmb{\theta}^\ast) \mathcal{F}(\pmb{\theta}^\ast)=\mathbf{0}$. Indeed, the conformation $\pmb{\theta}^\ast$  is an isolated and locally asymptotically stable equilibrium point for $\dot{\pmb{\theta}}=\mathcal{J}^\top(\pmb{\theta}) \mathcal{F}(\pmb{\theta})$ (see, e.g.,~\cite{tavousi2015protofold,kazerounian2005protofold,mohammadi2021quadratic}). In light of the properties of a folded conformation $\pmb{\theta}^\ast$, we can state our control objective  for unfolding as follows. 

\noindent\textbf{Protein Unfolding Control Problem (PUCP).} Consider the KCM-based protein conformation 
dynamics given by~\eqref{eq:KCM_ode} and a stable folded conformation given by $\pmb{\theta}^\ast$. Find a 
 closed-loop feedback control input $\mathbf{u}_c=\mathbf{u}_{\text{unfold}}(\pmb{\theta})$ for~\eqref{eq:KCM_ode} such that the folded conformation $\pmb{\theta}^\ast$ becomes an  unstable equilibrium for 
\begin{equation}\label{eq:KCM_odeUnfold}
\dot{\pmb{\theta}} = \mathbf{f}^{\text{unfold}}(\pmb{\theta}):= \mathcal{J}^\top(\pmb{\theta}) \mathcal{F}(\pmb{\theta}) + \mathbf{u}_{\text{unfold}}(\pmb{\theta}).
\end{equation} 

A \emph{trivial solution} to PUCP is given as follows. 

\begin{proposition}
	Consider $u_\text{unfold}(\pmb{\theta})=-2\mathcal{J}^\top(\pmb{\theta})\mathcal{F}(\pmb{\theta})$ for~\eqref{eq:KCM_odeUnfold}. Under this 
	control input the folded conformation $\pmb{\theta}^\ast$ will become a purely repulsing equilibrium for the resulting closed-loop dynamics given by $\dot{\pmb{\theta}}=-\mathcal{J}^\top(\pmb{\theta}) \mathcal{F}(\pmb{\theta})$. 
\end{proposition}

\begin{proof}
The folded conformation $\pmb{\theta}^\ast$  is an isolated and locally asymptotically stable equilibrium point for $\dot{\pmb{\theta}}=\mathcal{J}^\top(\pmb{\theta}) \mathcal{F}(\pmb{\theta})$. Therefore,  we can invoke  a converse argument~\cite{lin1996smooth} to conclude that there exists a Lyapunov function 
\begin{equation}
V^{\text{fold}}: \mathbb{R}^{2N} \to \mathbb{R}_+,\, \pmb{\theta} \mapsto  V^{\text{fold}}(\pmb{\theta})
\end{equation}
for these dynamics from which the local asymptotic stability of $\pmb{\theta}^\ast$ follows. It can be easily seen that 
$V^{\text{fold}}(\cdot)$ is indeed an anti-control Lyapunov function (ALF)~\cite{efimov2011oscillating} for $\dot{\pmb{\theta}}=-\mathcal{J}^\top(\pmb{\theta}) \mathcal{F}(\pmb{\theta})$,  where   $V^{\text{fold}}(\pmb{\theta}^\ast)=0$, $V^{\text{fold}}(\pmb{\theta})>0$ on $\mathcal{N}_{\pmb{\theta}^\ast}\backslash \{\pmb{\theta}^\ast\}$, where $\mathcal{N}_{\pmb{\theta}^\ast}$ is an open neighborhood of $\pmb{\theta}^\ast$, and   
	 $-\tfrac{\partial V^{\text{fold}}}{\partial \pmb{\theta}} \mathcal{J}^\top(\pmb{\theta}) \mathcal{F}(\pmb{\theta})>0$ for all $\pmb{\theta}\neq \pmb{\theta}^\ast$ on an open neighborhood of $\pmb{\theta}^\ast$ that is a subset of $\mathcal{N}_{\pmb{\theta}^\ast}$.
 \end{proof}
\begin{remark}
	There are some drawbacks associated with considering $V^{\text{fold}}(\cdot)$ as a candidate Chetaev function for protein unfolding. First, no closed-form expression for $V^{\text{fold}}(\cdot)$ is readily available. Second, $V^{\text{fold}}(\cdot)$  does not inherently possess the notion of direction of the unfolding forces. 
\end{remark}
\subsection{\textcolor{black}{Instability of folded conformations under optical tweezers}}
\label{subsec:analysispucp}
\textcolor{black}{In this section, we present an analysis to guarantee that the forces due to optical tweezers can be considered 
as solutions to PUCP formulated in Section~\ref{subsec:pucp}. Such  an analysis provides numerical conditions for real-time control of optical tweezer-based protein unfolding.} Using the torque in~\eqref{eq:unfoldTorque}, the unfolding conformation dynamics in~\eqref{eq:KCM_odeUnfold} become 
\begin{equation}
\dot{\pmb{\theta}} =  \mathbf{f}^\text{unfold}(\pmb{\theta}),  
\label{eq:mainDyn}
\end{equation} 
where $\mathbf{f}^\text{unfold}(\pmb{\theta}):=\mathcal{J}^\top(\pmb{\theta}) \big\{\mathcal{F}(\pmb{\theta}) + \mathcal{F}_\text{twz}(\pmb{\theta}) \big\}$. \textcolor{black}{Accordingly, if the protein molecule is under the effect of the forces generated by an optical tweezer as 
	in~\eqref{eq:optTwzForce}, then the unfolding dynamics are given by~\eqref{eq:KCM_odeUnfold}, where the unfolding control torque vector $\mathbf{u}_\text{unfold}(\pmb{\theta})$ is given by~\eqref{eq:unfoldTorque} and~\eqref{eq:genForce}.} Since the trap stiffness is modulated such that $\kappa(\pmb{\theta}^\ast)=0$ (see Remark~\ref{rem:modul}), the conformation $\pmb{\theta}^\ast$ is an equilibrium of~\eqref{eq:mainDyn}. 
%

\textcolor{black}{To analyze a folded conformation instability under
the effect
	of optical tweezers,} we are interested in deriving \textcolor{black}{numerical} conditions guaranteeing that PUCP objectives are met for the
unfolding dynamics in~\eqref{eq:mainDyn}. It is natural to expect that these conditions should depend on the direction of the 
applied force in~\eqref{eq:optTwzForce}. Considering the fact that a protein 
molecule extends under an optical tweezer-based force from its folded conformation 
$\pmb{\theta}^\ast$, we consider the candidate CF
\begin{equation}
\begin{aligned}
C_\text{twz}(\pmb{\theta}) = |\Delta \mathbf{r}_\text{NC}(\pmb{\theta},\pmb{\theta}^\ast)|^2 |\mathbf{r}_\text{NC}(\pmb{\theta}^\ast)|^2 \cos^2(\alpha_C) - \\
\big(\Delta \mathbf{r}_\text{NC}(\pmb{\theta},\pmb{\theta}^\ast)^\top \mathbf{r}_\text{NC}(\pmb{\theta}^\ast)\big)^2 , 
\label{eq:candidChetaev}
\end{aligned}
\end{equation}
for the unfolding dynamics in~\eqref{eq:mainDyn}. At any conformation $\pmb{\theta}$, the vector $\mathbf{r}_\text{NC}(\pmb{\theta})\in \mathbb{R}^3$ in~\eqref{eq:candidChetaev}
connects the nitrogen atom in the $\text{N}$-terminus to the carbon atom in the $\text{C}$-terminus.  Furthermore, the difference vector is given by $\Delta \mathbf{r}_\text{NC}(\pmb{\theta},\pmb{\theta}^\ast):=\mathbf{r}_\text{NC}(\pmb{\theta})-\mathbf{r}_\text{NC}(\pmb{\theta}^\ast)$. Finally, the constant $\alpha_C$ is any angle in-between $0$ and $\tfrac{\pi}{2}$. \textcolor{black}{Details on the geometric interpretation of this candidate CF is provided in Supplementary Material. As it will be seen in the next section (Propositions~\ref{prop:CCF} and \ref{prop:CCF2}), this candidate CF is also a CCF.} The proof of the next lemma \textcolor{black}{is given in Supplementary Material}. 
\begin{lemma}
	Consider the candidate CF $C_\text{twz}: \mathbb{R}^{2N}\to \mathbb{R}$ in~\eqref{eq:candidChetaev}. We have  $C_\text{twz}(\pmb{\theta}^\ast)=0$. Moreover,  there exists $\epsilon_0>0$ such that $\mathcal{C}_\text{twz}^{+}  \cap \mathcal{B}_{\pmb{\theta}^\ast}(\epsilon) \neq \emptyset$, where $\mathcal{C}_\text{twz}^{+}:= \big\{\pmb{\theta} \big|\, C_\text{twz}(\pmb{\theta}) > 0 \big\}$, for 
	all $\epsilon \in (0, \epsilon_0]$. 
	\label{lem:lemChet} 
\end{lemma}

Lemma~\ref{lem:lemChet} is used in the following proposition \textcolor{black}{whose proof is given in Supplementary Material}.

\begin{proposition}
	Consider the optical tweezer-based unfolding dynamics in~\eqref{eq:mainDyn}. The folded conformation $\pmb{\theta}^\ast$ is an unstable equilibrium for~\eqref{eq:mainDyn} if 
	%
	%
	\begin{equation}
	P(\pmb{\theta}):=\frac{\partial C_\text{twz}}{\partial \pmb{\theta}} \mathbf{f}^\text{unfold}(\pmb{\theta}) > 0 \text{ for all } \pmb{\theta}\in \mathcal{C}_\text{twz}^{+},
	\label{eq:CtwzCond}
	\end{equation} 
	\noindent where $\mathcal{C}_\text{twz}^{+}$ is defined in Lemma~\ref{lem:lemChet}, and
	\begin{equation}
	\begin{aligned}
	\frac{\partial C_\text{twz}}{\partial \pmb{\theta}} = 2 \Delta \mathbf{r}_{_\text{NC}}(\pmb{\theta},\pmb{\theta}^\ast)^\top \mathbf{M}(\pmb{\theta}^\ast)
	\frac{\partial \mathbf{r}_\text{NC}}{\partial \pmb{\theta}}, 
	\end{aligned}
	\label{eq:jacobChet}
	\end{equation} 
	where $\mathbf{M}(\pmb{\theta}^\ast) := |\mathbf{r}_{_\text{NC}}(\pmb{\theta}^\ast)|^2 \cos^2(\alpha_c) \mathbf{I}_3 - \mathbf{r}_{_\text{NC}}(\pmb{\theta}^\ast) \mathbf{r}_{_\text{NC}}(\pmb{\theta}^\ast)^\top$. 
	\label{prop:propUnfold}
\end{proposition}
%
%
%
\hspace{1ex} The condition given by~\eqref{eq:CtwzCond} can be \textcolor{black}{numerically} checked by investigating the Hessian of the function $P(\cdot)$ at the 
folded conformation under study as follows. 
\begin{proposition}
	Consider the folded protein molecule conformation $\pmb{\theta}^\ast$ for the closed-loop dynamics in~\eqref{eq:mainDyn} and the set 
	$\mathcal{C}_\text{twz}^{+}$ in Lemma~\ref{lem:lemChet}. Considering the Hessian matrix $\text{Hess}(P)|_{\pmb{\theta}^\ast}$, the condition given by~\eqref{eq:CtwzCond} holds if and only if  $\delta\pmb{\theta}^\top \text{Hess}(P)|_{\pmb{\theta}^\ast} \delta\pmb{\theta} > 0$ for all $\delta\pmb{\theta}$ such that $\pmb{\theta}^\ast + \delta\pmb{\theta} \in \mathcal{C}_\text{twz}^{+} \cap \mathcal{B}_{\pmb{\theta}^\ast}(\epsilon)$ for some $\epsilon > 0$.
	\label{prop:hessUnfold}
\end{proposition}
%
\begin{remark}
\textcolor{black}{The condition in Proposition~\ref{prop:hessUnfold}, which relies on numerical computation of $\text{Hess}(P)$ at the folded conformation $\pmb{\theta}^\ast$, can be used for real-time control of optical tweezers and AFMs in protein unfolding applications. In particular, to approximate each entry $\tfrac{\partial^2 P}{\partial \theta_i \partial \theta_j}$, one can use
	\begin{equation}
	\tfrac{P\big( \pmb{\theta}^\ast + \varepsilon_i \mathbf{1}_i + \varepsilon_j \mathbf{1}_j \big) - P\big( \pmb{\theta}^\ast + \varepsilon \mathbf{1}_i \big)-P\big( \pmb{\theta}^\ast + \varepsilon \mathbf{1}_j \big) + P( \pmb{\theta}^\ast)}{\varepsilon_i \varepsilon_j},
	\label{eq:HessianNum}
	\end{equation}
	where $P(\pmb{\theta}^\ast)=0$ because $\mathbf{f}^{\text{unfold}}(\pmb{\theta}^\ast)=\mathbf{0}$ (see Supplementary Material for analytical expression of these entries). Moreover, $\varepsilon_i$, $\varepsilon_j$  are some small positive constants and the vectors $\mathbf{1}_i$, $\mathbf{1}_j$ are the $i$-th, $j$-th columns of the identity matrix $\mathbf{I}_{2N}$, respectively (see, e.g.,~\cite{dennis1996numerical}).} \textcolor{black}{For real-time implementation, which is beyond the scope of the current article, there is a need for high-speed scanning and high-bandwidth control at nanoscale, where specialized DSP/FPGA-based embedded hardware have been developed (see, e.g.,~\cite{sun2010field,ragazzon2018model}).} 
\end{remark}
%
\subsection{\textcolor{black}{Synthesis of destabilizing control inputs for unfolding}}
\label{subsec:synthesispucp}
\textcolor{black}{In this section we present a class of CCFs for the KCM-based molecular dynamics. We then utilize these CCFs  in the  Artstein-Sontag universal formula, which was first presented in~\cite{efimov2014necessary} in the context of CCFs, to synthesize destabilizing state feedback control torques for  folded protein  conformations.  The following propositions provide two types of CCFs for the KCM-based dynamics of protein molecules, where the CCF in the second proposition is the candidate CF given by~\eqref{eq:candidChetaev}.} 
\begin{proposition}
\textcolor{black}{Consider the KCM-based dynamics in~\eqref{eq:KCM_ode} and a folded conformation $\pmb{\theta}^\ast$. 
	Consider the function $C: \mathcal{B}_{\pmb{\theta}^\ast}(\epsilon_0) \to \mathbb{R}$  given by
	\begin{equation}
	\begin{aligned}
	C(\pmb{\theta}) = g(|\Delta \mathbf{r}_\text{NC}(\pmb{\theta},\pmb{\theta}^\ast)|^2), 
	\end{aligned}
	\label{eq:candCCF}
	\end{equation}
	where $g: \mathbb{R}_{+} \to \mathbb{R}_{+}$ is any smooth strictly monotonic function such that $g(0)=0$. The function $C(\cdot)$ is a CCF for the nonlinear control-affine system in~\eqref{eq:KCM_ode}.}  
	\label{prop:CCF} 
\end{proposition}

\begin{proof}
	\textcolor{black}{Consider the set $\mathcal{C}^+ :=\{ \pmb{\theta}\in \mathcal{B}_{\pmb{\theta}^\ast}(\epsilon_0) : C(\pmb{\theta})>0 \}$. From~\eqref{eq:candCCF}, we have $C(\pmb{\theta}^\ast)=0$. Also, $\mathcal{C}^+ \cap \mathcal{B}_{\pmb{\theta}^\ast}(\epsilon)\neq \emptyset$  for all $\epsilon\in (0,\epsilon_0]$. Furthermore, 
$
\tfrac{\partial C}{\partial \pmb{\theta}} = g^\prime(|\Delta \mathbf{r}_\text{NC}(\pmb{\theta},\pmb{\theta}^\ast)|^2)\Delta \mathbf{r}_\text{NC}(\pmb{\theta},\pmb{\theta}^\ast)^\top \tfrac{\partial\mathbf{r}_\text{NC}}{\partial \pmb{\theta}}, 
$ where $g^{\prime}(.)$ is the derivative of $g(\cdot)$. We claim that 
\begin{equation}
\partial \mathcal{C}^{+} \cap \mathcal{C}^+ \cap \mathcal{B}_{\pmb{\theta}^\ast}(\epsilon) = \emptyset, 
\label{eq:CCFclaim}
\end{equation}
where  $\partial \mathcal{C}^{+} := \{\pmb{\theta} : \big|\tfrac{\partial C}{\partial \pmb{\theta}}\big| = 0\}$. To see this, note that 
\begin{equation}
\pmb{\theta} \in \partial\mathcal{C}^{+} \text{ if and only if } \Delta \mathbf{r}_\text{NC}(\pmb{\theta},\pmb{\theta}^\ast)^\top \tfrac{\partial\mathbf{r}_\text{NC}}{\partial \pmb{\theta}}=\mathbf{0}. 
\label{eq:borderEq}
\end{equation}
On the set $\mathcal{C}^+$, $\Delta \mathbf{r}_\text{NC}(\pmb{\theta},\pmb{\theta}^\ast)\neq \mathbf{0}$. Therefore, the relation in~\eqref{eq:borderEq} holds if and only if $\Delta \mathbf{r}_\text{NC}(\pmb{\theta},\pmb{\theta}^\ast)^\top\tfrac{\partial\mathbf{r}_\text{NC}}{\partial \theta_j}=\mathbf{0}$ for all 
$1 \leq j  \leq 2N$. Due to the kinematic structure of the protein molecule, which can be classified as a manipulator with  
hyper degrees of freedom (see~\cite{mochiyama1999shape}), we have    
\begin{equation}
\begin{aligned}
\frac{\partial\mathbf{r}_\text{NC}}{\partial \theta_j} =  \Xi(\pmb{\theta},\mathbf{u}_j^0) \mathbf{u}_j^0 \times (\mathbf{r}_\text{NC}(\pmb{\theta}) - \mathbf{r}_j(\pmb{\theta})),\, 1\leq j\leq 2N, 
\end{aligned}
\label{eq:CCF_struct}
\end{equation}
where $\Xi$ is given by~\eqref{eq:kineProtMat} and $\mathbf{r}_j(\pmb{\theta})$ is the position vector of the backbone chain atom ($\text{C}_\alpha$ atom for even $j$ and $N$ atom for odd $j$) in the $j$-th peptide plane.  Therefore, the relation in~\eqref{eq:borderEq} holds if and only if the nonzero vector $\Delta \mathbf{r}_\text{NC}(\pmb{\theta},\pmb{\theta}^\ast)$ is perpendicular to the collection of $2N$ three-dimensional vectors $\frac{\partial\mathbf{r}_\text{NC}}{\partial \theta_j}$. This geometric relation holds only if the protein molecule conformation $\pmb{\theta}$ is such that all the amino-acid linkages are co-located on the same  two-dimensional plane corresponding to an unfolded structure, which is impossible in a neighborhood of a folded conformation  $\pmb{\theta}^\ast$. Due to~\eqref{eq:CCFclaim}, we conclude that $\underset{\mathbf{u}_\text{c}\in \mathbb{R}^{2N}}{\sup}\{ \tfrac{\partial C}{\partial \pmb{\theta}} (\mathcal{J}^\top(\pmb{\theta}) \mathcal{F}(\pmb{\theta})+\mathbf{u}_\text{c})\}>0$ for all $\pmb{\theta}\in \mathcal{C}^+$. Therefore, $C(\cdot)$ is a CCF for~\eqref{eq:KCM_ode}.}  
\end{proof}
\begin{proposition}
	\textcolor{black}{Consider the KCM-based dynamics in~\eqref{eq:KCM_ode} and a folded conformation $\pmb{\theta}^\ast$. The function $C_{\text{twz}}(\cdot)$  in~\eqref{eq:candidChetaev} is a CCF for~\eqref{eq:KCM_ode}.} 
	%
\label{prop:CCF2}
\end{proposition}
\hspace{1ex} \textcolor{black}{Having a family of CCFs afforded by Propositions~\ref{prop:CCF} and~\ref{prop:CCF2}, we can use the following CCF-based Artstein-Sontag universal formula from~\cite{efimov2014necessary} 
\begin{equation}
\mathbf{u}_{\text{c}}(\pmb{\theta}) = -\phi\big[a(\pmb{\theta}),|B(\pmb{\theta})|\big] B(\pmb{\theta}), 
\label{eq:sontagInput}
\end{equation}
to synthesize unfolding control inputs that solve the PUCP formulated in Section~\ref{subsec:pucp}, where $a(\pmb{\theta}):=\tfrac{\partial C}{\partial \pmb{\theta}} \mathcal{J}^\top(\pmb{\theta}) \mathcal{F}(\pmb{\theta})$, $B^\top(\pmb{\theta}) := \tfrac{\partial C}{\partial \pmb{\theta}}$, and  $\phi(a,b):=\tfrac{a-\sqrt[p]{|a|^p+b^{2q}}}{b^2}$ if $b \neq 0$, and $\phi(a,b):=0$ if $b=0$, with any $2q\geq p > 1$, and $q>1$.} 
     \section{Simulation Results}
\label{sec:sims}
In this section we present simulation results to validate 
our proposed approach to protein unfolding using Chetaev instability framework. We consider a protein molecule chain consisting of $N-1=10$ peptide planes  with a $22$-dimensional dihedral angle space in~\eqref{eq:KCM_ode}. Our implementation follows the PROTOFOLD~I guidelines~\cite{kazerounian2005protofold,Tavousi2016}.

\begin{figure}[t]
	\centering
	\includegraphics[width=0.49\textwidth]{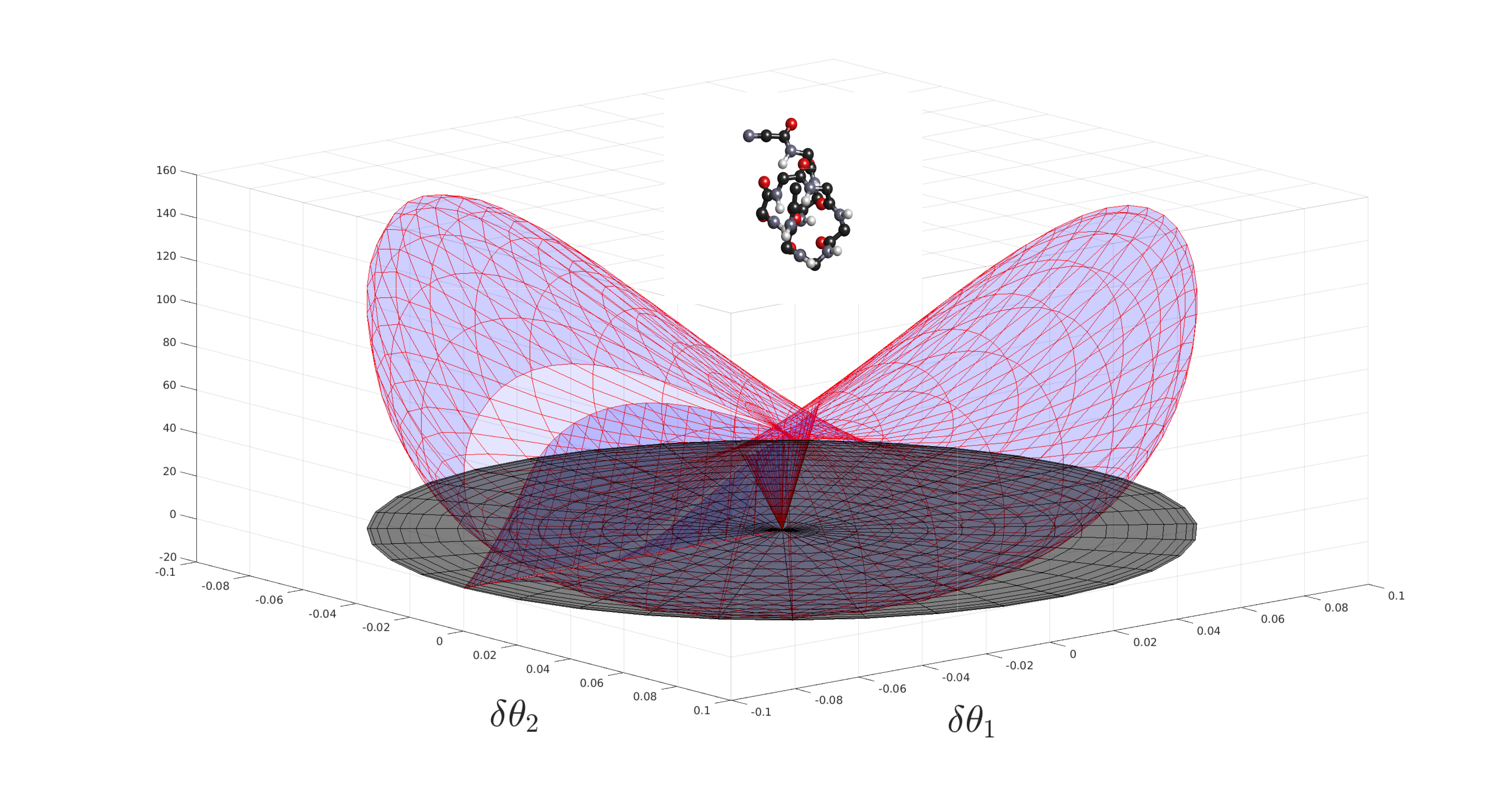}
	\caption{\small Visualization of the candidate Chetaev function $C_\text{twz}:\mathbb{R}^{22} \to \mathbb{R}$ based on the depicted 
		folded conformation at $\pmb{\theta}^\ast$.}
	\vspace{-3.0ex}
	\label{fig:chetProj}
\end{figure}

We have performed the unfolding simulations on the folded protein conformation $\pmb{\theta}^\ast$\footnote{$\pmb{\theta}^\ast=$ [1.34 1.37 1.26 1.29 1.42 1.73 1.65 1.65 1.46 1.62 1.49 2.01 1.31 0.99 1.98 1.9 1.59 1.56 1.57 0.93 1.22 1.29]$^\top$ in radians.} that has been obtained using the KCM-based iteration in~\cite{kazerounian2005protofold}. The candidate Chetaev function $C_\text{twz}:\mathbb{R}^{22} \to \mathbb{R}_{+}$ is chosen according to~\eqref{eq:candidChetaev} with $\alpha_C=\tfrac{\pi}{4}$. To visualize this function, we have plotted it using the dihedral angle vectors of the form $\pmb{\theta}=\pmb{\theta}^\ast+[\delta \theta_1,\delta \theta_2,\delta \theta_3, 0, \cdots, 0]^\top$, where the triplet $(\delta \theta_1,\delta \theta_2,\delta \theta_3)$ belong to the three-dimensional sphere of radius $0.1$ radians and centered at the origin. Figure~\ref{fig:chetProj} depicts both the folded conformation at $\pmb{\theta}^\ast$ and the visualization of the candidate Chetaev function when translated to the origin.  
%

\begin{figure}[t]
	\centering
	\includegraphics[width=0.37\textwidth]{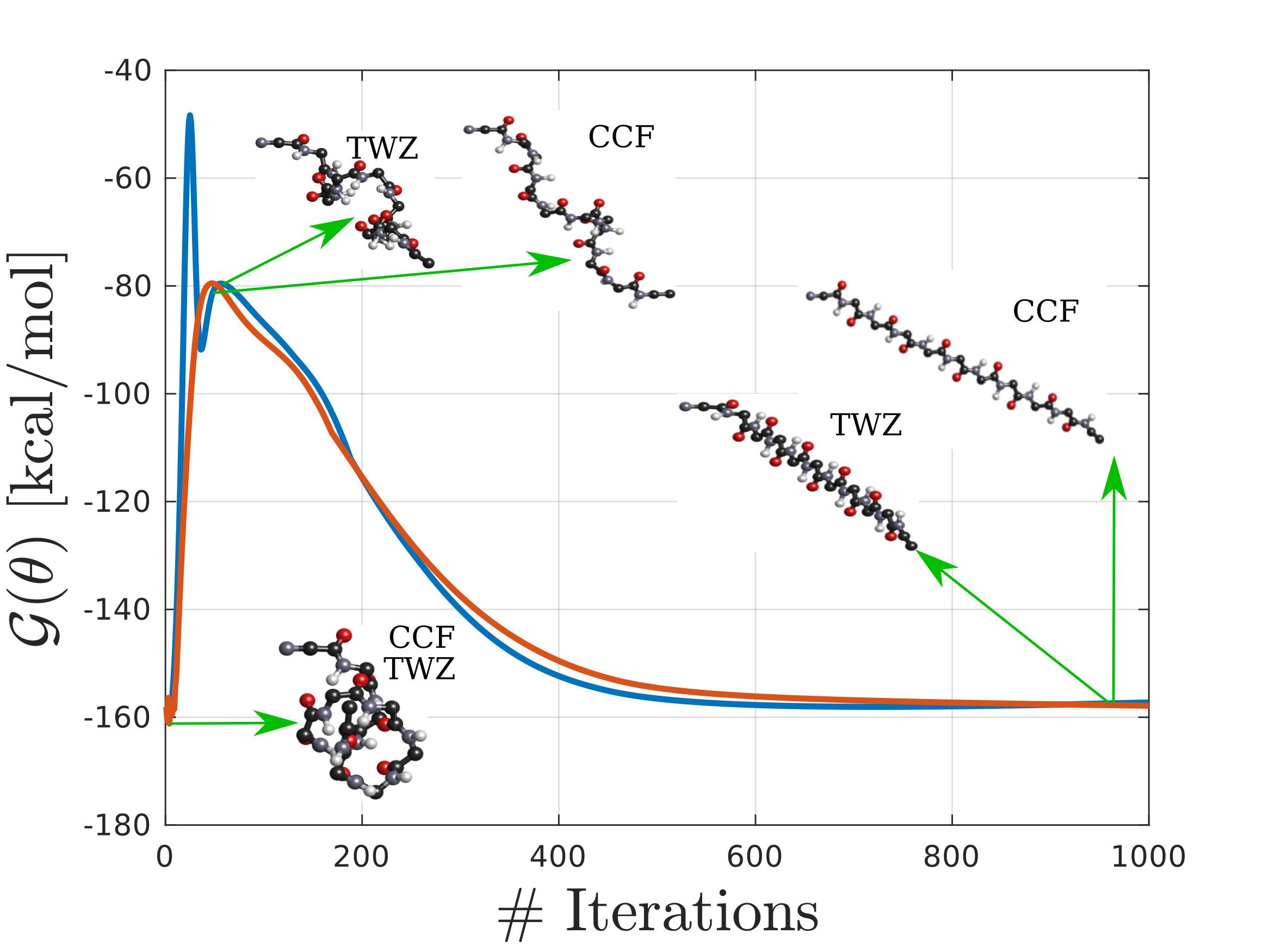}
	\caption{\small Energy profiles from unfolding numerical simulations under optical tweezer-based control 
		inputs  (red curve; molecule labeled by TWZ) and a theoretically-motivated input generated by  the 
		Artstein-Sontag universal formula (blue curve; molecule labeled by CCF).}
	\vspace{-3.0ex}
	\label{fig:unfoldSim}
\end{figure}

Using the setup above, we conduct two numerical simulations. The first one entails the optical tweezer-based unfolding while the second one entails an unfolding simulation using the Artstein-Sontag universal formula in~\eqref{eq:sontagInput} with $p=q=2$ and  $C_\text{twz}$ in Proposition~\ref{prop:CCF2} as the CCF. In the first simulation, where the numerical condition of Proposition~\ref{prop:hessUnfold} is verified, we let the optical tweezer displacement be $\mathbf{x}_{\text{twz}}=x_0 \tfrac{\mathbf{r}_\text{NC}(\pmb{\theta}^\ast)}{\|\mathbf{r}_\text{NC}(\pmb{\theta}^\ast)\|}$, where $x_0$ is chosen to be $51$ nm. In other words, we elongated the protein molecule along the vector that connects the $N$-terminus to the $C$-terminus.   Furthermore, we utilized a modulated optical trap stiffness of $\kappa(\pmb{\theta})=\kappa_0|\mathbf{r}_\text{NC}(\pmb{\theta}^\ast) - \mathbf{r}_\text{NC}(\pmb{\theta})|^2$, where $\kappa_0=0.16 \text{ pN/nm}$. Figure~\ref{fig:unfoldSim} depicts the free energy of the molecule under the optical tweezer-based and the CCF-based unfolding control inputs  as well as several protein conformations along their respective  unfolding pathways (see Supplementary Material for further discussion). 
    \section{Conclusion}
\label{sec:conc}
This paper formulated the important biochemical process of KCM-based protein unfolding as a destabilizing control analysis/synthesis problem. In light of this formulation and using the Chetaev instability framework, numerical conditions on optical tweezer forces that lead to protein unfolding have been derived. Furthermore, after presenting two types of CCFs, a theoretically-motivated CCF-based Artstein-Sontag unfolding input has been derived. These findings lead us to further research avenues such as encoding entropy-loss constraints for protein unfolding using the obtained CCFs, investigating folding pathways for two-state proteins, and studying protein folding/unfolding in solutions.
%
%
     
     


\vspace{-1ex}
\bibliographystyle{./styles/IEEEtran}
\bibliography{references}

\clearpage
\newcounter{defcounter}
\setcounter{defcounter}{0}
\newenvironment{myequation}{%
	\addtocounter{equation}{-1}
	\refstepcounter{defcounter}
	\renewcommand\theequation{s\thedefcounter}
	\begin{equation}}
{\end{equation}}
\renewcommand\thefigure{SF\arabic{figure}}
\setcounter{figure}{0}   
\section*{Supplementary Material To ``Chetaev Instability Framework for Kinetostatic Compliance-Based Protein Unfolding''}

\noindent{\textbf{Notation.}} In these supplementary notes, we will number the equations using (s\#), the references using [S\#], and the figures using SF\#. Therefore, (s1) refers to Equation~1 in the supplementary notes while (1) refers to Equation~1 in the original article. The same holds true for the references and the figures. 

\subsection{Schematic of protein folding/unfolding against the free energy landscape} 

\begin{figure}[h]
	\centering
	\includegraphics[width=0.42\textwidth]{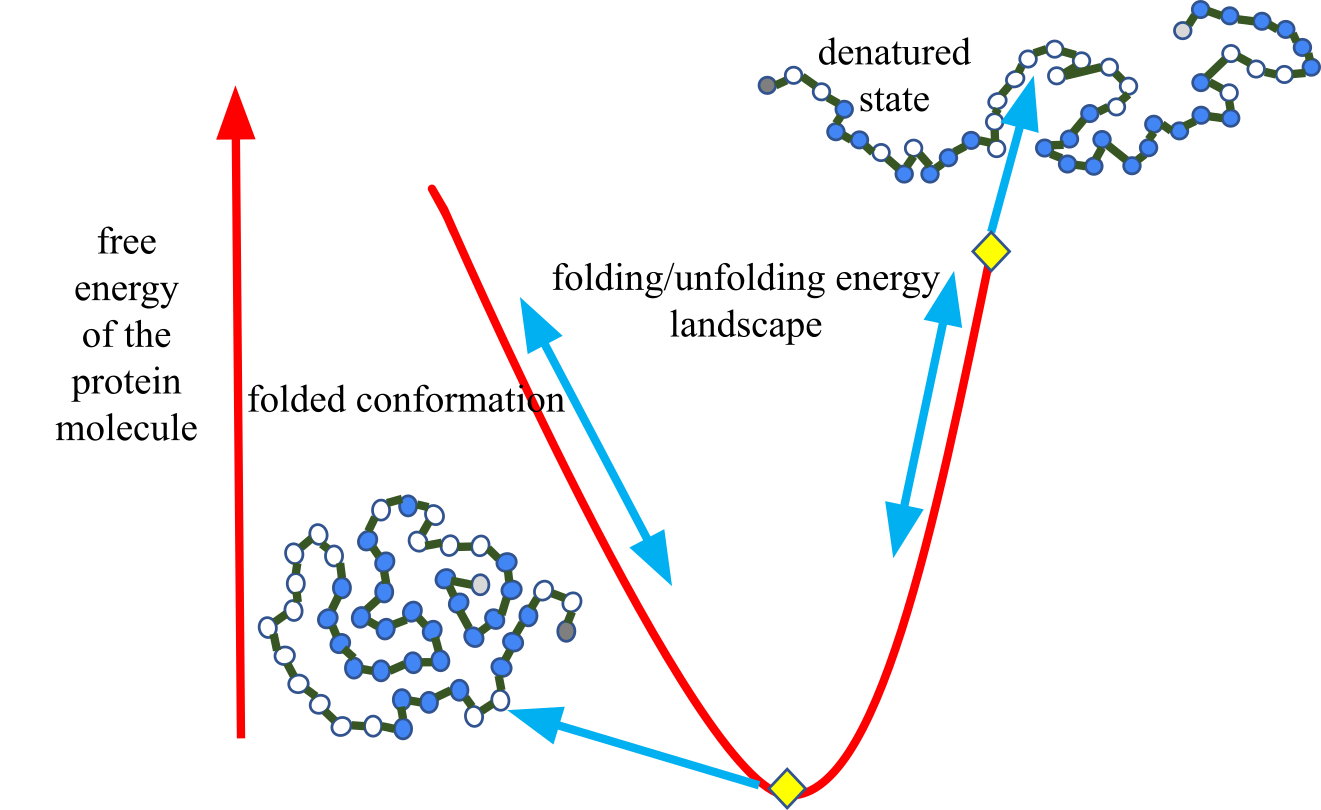}
	\caption{\small Protein folding/unfolding against the free energy landscape of the molecule.}
	\vspace{-3ex}
	\label{fig:foldingSchem}
\end{figure}

Figure~\ref{fig:foldingSchem} depicts a schematic of folding/unfolding against the free energy landscape of the protein molecule. 

\subsection{Principles of operation of optical tweezers}
\label{subsec:optTweezer}

\begin{figure}[h]
	\centering
	\includegraphics[scale=0.45]{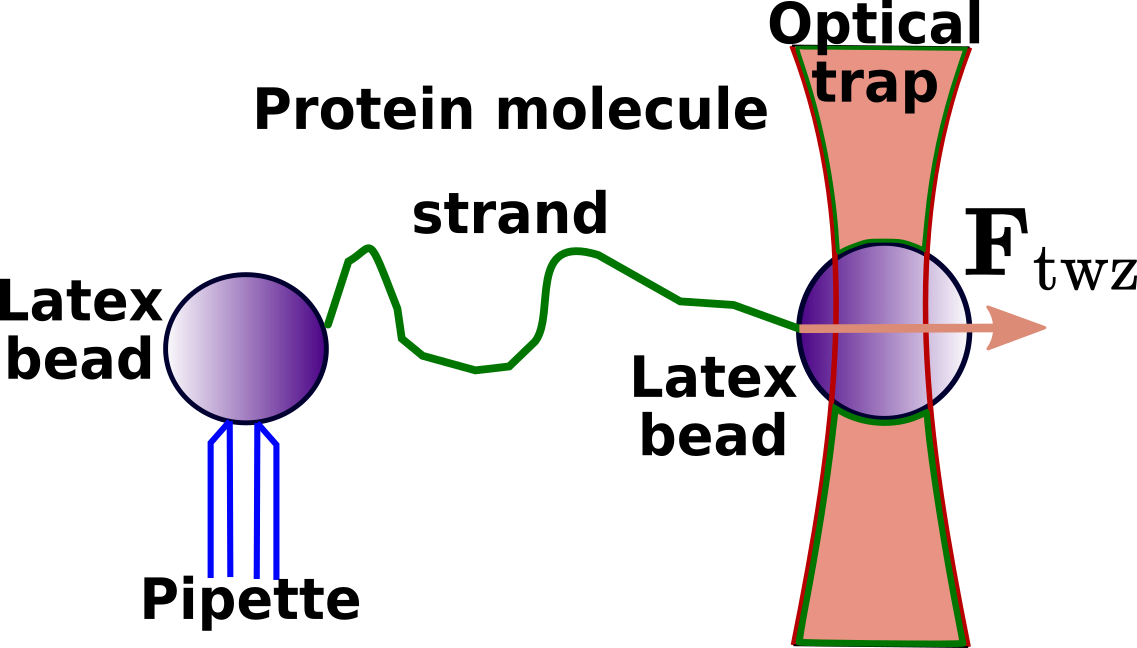}
	\caption{\small Schematic of optical tweezer-based protein unfolding (recreated from~[S1]).}
	\vspace{-6ex}
	\label{fig:optTwz}
\end{figure}

Optical tweezers, which were discovered and developed by Ashkin~[S2], operate based on the principle of force exertion by light on matter due to the change in  momentum that light carries. Figure~\ref{fig:optTwz} 
depicts a schematic of optical tweezer-based protein unfolding.  Optical tweezers can be utilized for unfolding protein molecules by chemically attaching a latex bead (microsphere) to the protein under study.  The bead could subsequently be trapped and utilized both as a handle for manipulating the protein molecule and as a tool for measuring displacements at a very small scale. Under the assumptions of a nearly Gaussian beam and a locally harmonic potential energy for a given optical tweezer, the optical trap can be modeled as a Hookean spring~\cite{bustamante2020single}.

\subsection{On the choice of the candidate CF}

\begin{figure}[h]
	\centering
	\includegraphics[width=0.25\textwidth]{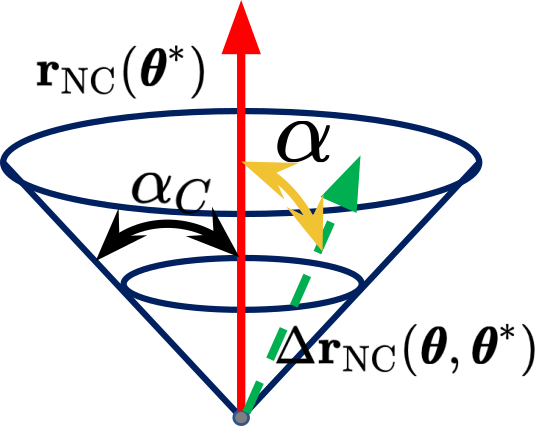}
	\caption{\small When $C_\text{twz}(\pmb{\theta})=0$, $\Delta \mathbf{r}_\text{NC}(\pmb{\theta},\pmb{\theta}^\ast)$ belongs to a right circular cone with vertex at the origin,  axis parallel to $\mathbf{r}_\text{NC}(\pmb{\theta}^\ast)$, and aperture $2\alpha_C$.} 
	\vspace{-2.5ex}
	\label{fig:coneChetaevGeom}
\end{figure}

Figure~\ref{fig:coneChetaevGeom} represents the geometric interpretation associated with the three dimensional vectors utilized for defining $C_\text{twz}(\cdot)$ in~\eqref{eq:candidChetaev}. \textcolor{black}{The vectors $\mathbf{r}_\text{NC}(\pmb{\theta})$ and $\mathbf{r}_\text{NC}(\pmb{\theta}^{\ast})$ represent the ``length'' of the molecule at a general conformation $\pmb{\theta}$ and the folded conformation  $\pmb{\theta}^{\ast}$, respectively. The positive level sets of the 
candidate CF $C_\text{twz}(\cdot)$ in~\eqref{eq:candidChetaev}, which depends on $\mathbf{r}_\text{NC}(\pmb{\theta})$ and $\mathbf{r}_\text{NC}(\pmb{\theta}^{\ast})$, correspond to conformations at which the protein molecule has been elongated  from the folded conformation along desired directions.  Indeed, when $\dot{C}_\text{twz}>0$ along the trajectories of the unfolding dynamics on the positive level sets of $C_\text{twz}(\cdot)$, the molecule is being elongated away from its folded conformation.}

\subsection{Proof of Technical Results}

\noindent\textbf{Proof of Lemma~\ref{lem:lemChet}.} Since $\Delta \mathbf{r}_\text{NC}(\pmb{\theta}^\ast,\pmb{\theta}^\ast)=\mathbf{0}$, we conclude that $C_\text{twz}(\pmb{\theta}^\ast)=0$. Let us define the mapping 
	\begin{myequation} 
	\begin{aligned}
	f: \mathbb{R}^{2N}\backslash \{\pmb{\theta}^\ast\} \to \mathbb{R},\, \pmb{\theta} \mapsto \cos^{-1}(\tfrac{\Delta \mathbf{r}_\text{NC}(\pmb{\theta},\pmb{\theta}^\ast)^\top \mathbf{r}_\text{NC}(\pmb{\theta}^\ast)}{|\Delta \mathbf{r}_\text{NC}(\pmb{\theta},\pmb{\theta}^\ast)| |\mathbf{r}_\text{NC}(\pmb{\theta}^\ast)|}),  
	\end{aligned}
	\label{eq:fDefn}
	\end{myequation}
\hspace{-1.1ex}\textcolor{black}{where $f(\pmb{\theta})$ is the angle between $\mathbf{r}_\text{NC}(\pmb{\theta}^\ast)$ and $\Delta \mathbf{r}_\text{NC}(\pmb{\theta},\pmb{\theta}^\ast)$.} We claim 
	\begin{myequation}
	\mathcal{C}_\text{twz}^{+} = f^{-1}\big( (\alpha_C,\tfrac{\pi}{2}) \cap (-\tfrac{\pi}{2},-\alpha_C) \big).
	\label{eq:VplusDefn}  
	\end{myequation}
	Consider an arbitrary 
	$\pmb{\theta}_0 \in \mathbb{R}^{2N}$ such that $C_\text{twz}(\pmb{\theta}_0) > 0$. From~\eqref{eq:candidChetaev} and~\eqref{eq:fDefn}, it can be seen that 
	\begin{myequation}
	C_\text{twz}(\pmb{\theta}_0) = |\Delta \mathbf{r}_{_\text{NC}}(\pmb{\theta}_0,\pmb{\theta}^\ast)|^2 |\mathbf{r}_{_\text{NC}}(\pmb{\theta}^\ast)|^2 ( \cos^2(\alpha_C) \text{-} \cos^2(f(\pmb{\theta}_0))). 
	\end{myequation}
	Therefore, $C_\text{twz}(\pmb{\theta}_0) > 0$ if and only if $\pmb{\theta}_0 \in \mathcal{C}_\text{twz}^{+}$. Since $f(\cdot)$ is a 
	continuous function,  we conclude that $\mathcal{C}_\text{twz}^{+}$ in~\eqref{eq:VplusDefn}  is an open set and $\pmb{\theta}^\ast$ belongs 
	to the boundary of $\mathcal{C}_\text{twz}^{+}$. Therefore, $\mathcal{C}_\text{twz}^{+}  \cap \mathcal{B}_{\pmb{\theta}^\ast}(\epsilon) \neq \emptyset$ 
	for all $\epsilon \in (0, \epsilon_0]$ for some $\epsilon_0>0$.\hfill$\blacksquare$\\

\noindent\textbf{Proof of Proposition~\ref{prop:propUnfold}.} Considering the smooth candidate Chetaev function in~\eqref{eq:candidChetaev}, the directional derivative $D  C_\text{twz}(\pmb{\theta})\mathbf{f}^{\text{unfold}}(\pmb{\theta})$ in the direction of $\mathbf{f}^\text{unfold}(\cdot)$  is given by $\tfrac{\partial C_\text{twz}}{\partial \pmb{\theta}} \mathbf{f}^{\text{unfold}}(\pmb{\theta})$, where 
\begin{myequation}
	\begin{aligned}
	\frac{\partial C_\text{twz}}{\partial \pmb{\theta}} = 2 \bigg\{ |\mathbf{r}_{_\text{NC}}(\pmb{\theta}^\ast)|^2 \cos^2(\alpha_C)  \Delta \mathbf{r}_{_\text{NC}}(\pmb{\theta},\pmb{\theta}^\ast)^\top - \\ 
	\Delta \mathbf{r}_{_\text{NC}}(\pmb{\theta},\pmb{\theta}^\ast)^\top \mathbf{r}_{_\text{NC}}(\pmb{\theta}^\ast)\mathbf{r}_{_\text{NC}}(\pmb{\theta}^\ast)^\top \bigg\} \frac{\partial \mathbf{r}_\text{NC}}{\partial \pmb{\theta}}.
	\end{aligned}
\end{myequation}
Therefore, due to  Lemma~\ref{lem:lemChet}, the statement of the proposition follows from Theorem~\ref{thm:cheta}.\hfill$\blacksquare$\\

\noindent\textbf{Proof of Proposition~\ref{prop:hessUnfold}.} 	Given the $C^\infty$-smoothness of $P(\cdot)$ in a neighborhood of $\pmb{\theta}^\ast$, we can utilize the Taylor's theorem for multivariate functions to arrive at  
\begin{myequation}
P(\pmb{\theta}) = P(\pmb{\theta}^\ast) + \frac{\partial P}{\partial \pmb{\theta}}\bigg|_{\theta^\ast} \delta\pmb{\theta} + \delta\pmb{\theta}^\top \text{Hess}(P)|_{\pmb{\theta}^\ast} \delta\pmb{\theta} + \text{ H.O.T}, 
\end{myequation}
where $\pmb{\theta} \in \mathcal{B}_{\pmb{\theta}^\ast}(\epsilon) \cap \mathcal{V}^+$ for some $\epsilon > 0$ and $\delta \pmb{\theta} := \pmb{\theta} - \pmb{\theta}^\ast$. It can be seen that $P(\pmb{\theta}^\ast) = 0$ because $\pmb{\theta}^\ast$ is an equilibrium for the dynamics in~\eqref{eq:mainDyn}. Furthermore, $\tfrac{\partial P}{\partial \pmb{\theta}}|_{\pmb{\theta}^\ast}=\mathbf{0}$ because $\mathbf{f}^\text{unfold}(\pmb{\theta}^\ast)=\mathbf{0}$ and  $\tfrac{\partial C_\text{twz}}{\partial \pmb{\theta}}|_{\pmb{\theta}^\ast} =\mathbf{0}$ according to~\eqref{eq:jacobChet}. Therefore, $P(\pmb{\theta})>0$ on $\mathcal{B}_{\pmb{\theta}^\ast}(\epsilon) \cap \mathcal{V}^+$ if and only if the stated condition in the proposition holds.\hfill$\blacksquare$ \\ 

\noindent{\textbf{Proof of Proposition~\ref{prop:CCF2}}.} 
	The proof is almost verbatim the proof of Proposition~\ref{prop:CCF} with the difference that the relation in~\eqref{eq:borderEq2} should be 
	replaced with  
	\begin{myequation}
	\pmb{\theta} \in \partial\mathcal{C}^{+} \text{ if and only if } \Delta \mathbf{r}_\text{NC}(\pmb{\theta},\pmb{\theta}^\ast)^\top \mathbf{M}(\pmb{\theta}^\ast) \tfrac{\partial\mathbf{r}_\text{NC}}{\partial \pmb{\theta}}=\mathbf{0}. 
	\label{eq:borderEq2}
	\end{myequation}
	Since $\mathbf{M}(\pmb{\theta}^\ast)$ is a symmetric and non-singular matrix with $\det(\mathbf{M})=-\sin^2(\alpha_c) |\mathbf{r}_\text{NC}(\pmb{\theta}^\ast)|^6$, the relation in~\eqref{eq:borderEq2} holds if and only if the nonzero vector $\mathbf{M}(\pmb{\theta}^\ast)\Delta \mathbf{r}_\text{NC}(\pmb{\theta},\pmb{\theta}^\ast)$ is perpendicular to the collection of $2N$ three-dimensional vectors $\frac{\partial\mathbf{r}_\text{NC}}{\partial \theta_j}$. The rest of the proof follows verbatim the proof of 
	Proposition~\ref{prop:CCF}. \hfill$\blacksquare$ \\

\noindent{\textbf{The Analytical Expression for the Entries of Hessian}.} The entries of the 
Hessian of the function $P(\cdot)$ given by~\eqref{eq:CtwzCond} at a folded conformation $\pmb{\theta}^{\ast}$
take the form
\begin{myequation}
	\begin{aligned}
	\tfrac{\partial^2 P}{\partial \theta_j \partial \theta_k}\big|_{\pmb{\theta}^\ast} = & 2 \sum_{i=1}^{2N} \tfrac{\partial \mathbf{r}_\text{NC}}{\partial \theta_i}(\pmb{\theta}^\ast) \mathbf{M}(\pmb{\theta}^\ast)\big\{ \tfrac{\partial \mathbf{r}_{\text{NC}}(\pmb{\theta}^\ast)}{\partial \theta_j} \tfrac{\partial f_{i,\text{unfold}}(\pmb{\theta}^\ast)}{\partial \theta_k}  + \\ &
	\tfrac{\partial \mathbf{r}_{\text{NC}}(\pmb{\theta}^\ast)}{\partial \theta_k} \tfrac{\partial f_{i,\text{unfold}}(\pmb{\theta}^\ast)}{\partial \theta_j}\big\},\; 1 \leq j,\, k\leq 2N, 
    \end{aligned}
\end{myequation}
\hspace{-1ex}where $\mathbf{M}(\pmb{\theta}^\ast)$ is defined in Proposition~\ref{prop:propUnfold}, the vector $\mathbf{r}_\text{NC}(\pmb{\theta})\in \mathbb{R}^3$ connects the nitrogen atom in the $\text{N}$-terminus to the carbon atom in the $\text{C}$-terminus, and $f_{i,\text{unfold}}(\pmb{\theta})$ is the $i$-th element of the vector $\mathbf{f}^\text{unfold}(\pmb{\theta})$ in~\eqref{eq:mainDyn}.

\subsection{Control Chetaev Functions}

In this section we provide some preliminaries on control Chetaev functions due to Efimov and collaborators~[S3]. The reader is referred to the line of work by  Efimov, Kellett, and collaborators~[S3] and [S4] for further details. 

Consider the nonlinear control-affine system 
\begin{myequation}
\dot{\mathbf{x}}= \mathbf{F}(\mathbf{x}) + \mathbf{G}(\mathbf{x})\mathbf{u},
\label{eq:nonlinSys1}
\end{myequation}
\hspace{-1ex}where $\mathbf{x}\in\mathbb{R}^n$ is the state vector, $\mathbf{u}\in\mathbb{R}^m$ is the control input, $\mathbf{F}(\mathbf{x}^\ast)=\mathbf{0}$, and  $\mathbf{F}:\mathbb{R}^n \to \mathbb{R}^n$ and $\mathbf{G}:\mathbb{R}^m \to \mathbb{R}^n$ are locally Lipschitz continuous functions.  A smooth function $V:\mathcal{B}_{\mathbf{x}^\ast}(\epsilon)\to \mathbb{R}$, such that $V(\mathbf{x}^\ast)=0$ and $\mathcal{V}^+\cap \mathcal{B}_{\mathbf{x}^\ast}(\epsilon)\neq \emptyset$, where  $\mathcal{V}^+:=\{\mathbf{x}: V(\mathbf{x})> 0\}$ and for any $\epsilon\in (0,\epsilon_0]$,  is called a CCF for the control system in~\eqref{eq:nonlinSys1} if 
\begin{myequation}
\sup\limits_{\mathbf{u}\in \mathbb{R}^m} \{D^- V(\mathbf{x})\mathbf{F}(\mathbf{x}) + (D^- V(\mathbf{x})\mathbf{G}(\mathbf{x}))^\top \mathbf{u}\} > 0, 
\label{eq:CCF_V}
\end{myequation}
\hspace{-1ex}for all  $\mathbf{x}\in \mathcal{V}^+$. As it is argued in [S3], if for all $\mathbf{x}\in \mathcal{V}^+$ at which  
$|D^- V(\mathbf{x})\mathbf{G}(\mathbf{x})|=0$, it holds that $|D^- V(\mathbf{x})\mathbf{F}(\mathbf{x})|>0$, then $V(\cdot)$ is a CCF for~\eqref{eq:nonlinSys1}. Furthermore, in the special case when $\mathbf{G}(\mathbf{x})=\mathbf{I}_{n\times n}$ and $\mathbf{u}\in\mathbb{R}^n$, if $|D^- V(\mathbf{x})|\neq 0$ for all $\mathbf{x}\in \mathcal{V}^+$, then $V(\cdot)$ is a CCF for~\eqref{eq:nonlinSys1}. 

\section*{Supplementary References}
\small{
\noindent [S1]\hspace{1pt} J. L. Killian, F. Ye, and M. D. Wang, ``Optical tweezers: a force to be
reckoned with,'' \emph{Cell}, vol. 175, no. 6, pp. 1445–1448, 2018.\\

\noindent [S2]\hspace{1pt} A. Ashkin, ``Acceleration and trapping of particles by radiation pressure,'' \emph{Phys. Rev. 
Lett.}, vol. 24, no. 4, p. 156, 1970.\\

\noindent [S3]\hspace{1pt} D. Efimov, W. Perruquetti, and M. Petreczky, ``On necessary conditions
of instability and design of destabilizing controls,'' in \emph{53rd IEEE Conf.
Dec. Contr. (CDC).} IEEE, 2014, pp. 3915--3917.\\

\noindent [S4]\hspace{1pt} P. Braun, L. Gr\"{u}ne, and C. M. Kellett, ``Complete instability of ¨
differential inclusions using Lyapunov methods,'' in \emph{57th IEEE Conf.
Dec. Contr. (CDC).} IEEE, 2018, pp. 718--724.

}

\end{document}